\documentstyle[mprocl]{article}

\input{psfig}
\def\Journal#1#2#3#4{{#1} {\bf #2}, #3 (#4)}

\def\apj{\em ApJ}
\def\apjs{\em ApJS}

\def\mnras{\em MNRAS}

\def\bm{{\bf m}}
\def\bd{{\bf d}}

\def\hmpc{\,{\rm h^{-1}Mpc}}
\def\ihmpc{\, h\, {\rm Mpc^{-1}}}

\def\3hmpc{\, ( h^{-1} {\rm Mpc})^3}
\def\ln{{\rm ln}}
\def\lcdm{$\Lambda$CDM}   

\def\h5{h_{50}}
\def\P{{\cal P}} 
\def\L{{\cal L}} 

\def\be{\begin{equation}}
\def\ee{\end{equation}}
\def\bea{\begin{eqnarray}}
\def\eea{\end{eqnarray}}
\begin{document}

\title{LARGE-SCALE MASS POWER SPECTRUM FROM \\ PECULIAR VELOCITIES}

\author{ I. ZEHAVI }

\address{Racah Institute of Physics, The Hebrew University, Jerusalem 91904,
Israel}

\maketitle\abstracts{
This is a brief progress report on a long-term collaborative project
to measure the power spectrum (PS) of mass density fluctuations from
the Mark III and the SFI catalogs of peculiar
velocities.\cite{zz}$^{\!,\,}$\cite{fz} The PS is estimated by
applying maximum likelihood analysis, using generalized CDM models
with and without COBE normalization.  The application to both catalogs
yields fairly similar results for the PS, and the robust results are
presented. }
  
\section{Introduction}
\label{sec:intro}
In the standard picture of cosmology, structure evolved from small
density fluctuations that grew by gravitational instability. These
initial fluctuations are assumed to have a Gaussian distribution
characterized by the PS. On large scales, the fluctuations are linear
even at late times and still governed by the initial PS. The PS is
thus a useful statistic for large-scale structure, providing
constraints on cosmology and theories of structure formation. In
recent years, the galaxy PS has been estimated from several redshift
surveys.\cite{SW} \,
In this work, we develop and apply likelihood analysis\cite{k} in
order to estimate the {\it mass} PS from peculiar velocity
catalogs. Two such catalogs are used. One is the Mark III catalog of
peculiar velocities,\cite{mark3} a compilation of several data sets,
consisting of roughly 3000 spiral and elliptical galaxies within a
volume of $\sim 80 \hmpc$ around the local group, grouped into $\sim
1200$ objects. The other is the recently completed SFI
catalog,\cite{sfi} a homogeneously selected sample of $\sim 1300$
spiral field galaxies, which complies with well-defined criteria.  It
is interesting to compare the results of the two catalogs, especially
in view of apparent discrepancies in the appearance of the velocity
fields.\cite{sfiPOT}$^{\!,\,}$\cite{ITF}

\section{Method}
\label{sec:method}
Given a data set $\bd$, the goal is to estimate the most likely model
$\bm$. Invoking a Bayesian approach, this can be turned to maximizing
the likelihood function $\L \equiv \P(\bd|\bm)$, the probability of
the data given the model, as a function of the model parameters.
Under the assumption that both the underlying velocities and the
observational errors are Gaussian random fields, the likelihood
function can be written as $ {\cal L} = [ (2\pi)^N
\det(R)]^{-1/2} \exp\left( -{1\over 2}\sum_{i,j}^N {d_i R_{ij}^{-1}
d_j}\right), $ where $\{d_i\}_{i=1}^{N}$ is the set of observed peculiar
velocities and $R$ is their correlation matrix. $R$ involves the theoretical 
correlation, calculated in linear theory for each assumed cosmological
model, and the estimated covariance of the errors.

The likelihood analysis is performed by choosing some parametric
functional form for the PS. For each assumed PS, one can calculate the
likelihood function and going over the parameter space find the the PS
parameters that provide the maximum likelihood.  Confidence levels are
estimated by approximating $-2\ln\L$ as a $\chi^2$ distribution with
respect to the model parameters. Note that this method, based on
peculiar velocities, essentially measures $f(\Omega)^2P(k)$ and not
the mass density PS by itself. Careful testing of the method was done
using realistic mock catalogs,\cite{mockM3} designed to mimic in
detail the real catalogs.

We use several models for the PS. One of these is the so-called
$\Gamma$ model, where we vary the amplitude and the shape-parameter
$\Gamma$.  The main analysis is done with a suit of generalized CDM
models, normalized by the COBE 4-yr data. These include open models,
flat models with a cosmological constant and tilted models with or
without a tensor component.  The free parameters are then the density
parameter $\Omega$, the Hubble parameter $h$ and the power index $n$.
The recovered PS is sensitive to the assumed observational errors,
that go as well into $R$. We extend the method such that also the
magnitude of these errors is determined by the likelihood analysis, by
adding free parameters that govern a global change of the assumed
errors, in addition to modeling the PS. We find, for both catalogs, a
good agreement with the original error estimates, thus allowing for a
more reliable recovery of the PS.

\section{Results}
\label{sec:res}

\begin{figure}
\vskip -4.3cm
\psfig{figure=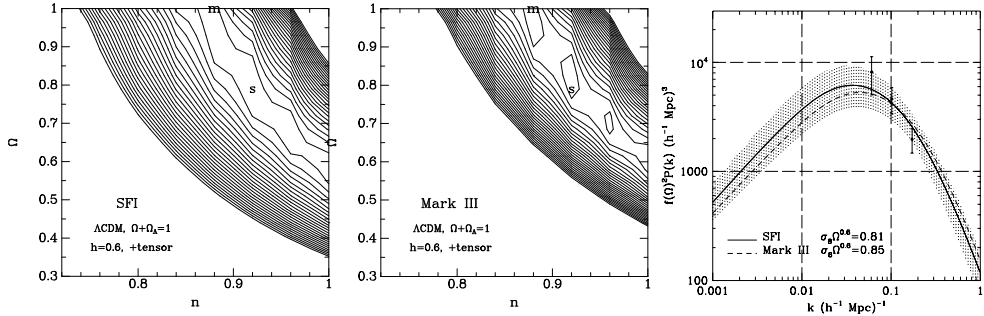,height=7.0in}
\vskip -9.6cm 
\caption{Likelihood analysis results for the flat \lcdm\ model with $h=0.6$.
$\ln\L$ contours in the $\Omega-n$ plane are shown for SFI (left
panel) and Mark III (middle). The best-fit parameters are marked by
`s' and `m' on both, for SFI and Mark III respectively. The right
panel shows the corresponding PS for the SFI case (solid line) and for
Mark III (dashed). The shaded region is the SFI $90\%$ confidence
region. The three dots are the PS calculated from Mark III by Kolatt
and Dekel (1997),$^{10}$ together with their $1\sigma$ error-bar.}
\label{fig}
\end{figure}

Figure~\ref{fig} shows, as a typical example, the results for the flat
\lcdm\ family of models, with a tensor component in the initial
fluctuations, when setting $h=0.6$ and varying $\Omega$ and $n$. The
left panel shows the $\ln\L$ contours for the SFI catalog and the
middle panel the results for Mark III. As can be seen from the
elongated contours, what is determined well is not a specific point
but a high likelihood ridge, constraining a degenerate combination of
the parameters of the form $\Omega\, n^{3.7} = 0.59 \pm 0.08$, in this
case. The right panel shows the corresponding maximum-likelihood PS
for the two catalogs, where the shaded region represents the $90\%$
confidence region obtained from the SFI high-likelihood ridge.

These results are representative for all other PS models we tried. For
each catalog, the different models yield similar best-fit PS, falling
well within each others formal uncertainties and agreeing especially
well on intermediate scales ($k \sim 0.1\ihmpc$). The similarity, seen
in the figure, of the PS obtained from SFI to that of Mark III is
illustrative for the other models as well. This indicates that the
peculiar velocities measured by the two data sets, with their
respective error estimates, are consistent with arising from the same
underlying mass density PS.  Note also the agreement with an
independent measure of the PS from the Mark III catalog, using the
smoothed density field recovered by POTENT (the three dots).\cite{kd}

The robust result, for both catalogs and all models, is a relatively
high PS, with $P(k) \Omega^{1.2} = (4.5\pm2.0)\times 10^3 \3hmpc$ at
$k=0.1\ihmpc$.  An extrapolation to smaller scales using the different
CDM models gives $\sigma_8 \Omega^{0.6} = 0.85 \pm 0.2$.  The
error-bars are crude, reflecting the $90\%$ formal likelihood
uncertainty for each model, the variance among different models and
between catalogs. The general constraint of the high likelihood ridges
is of the sort $\Omega\, {\h5}^\mu\, n^\nu = 0.75 \pm 0.25$, where
$\mu = 1.3$ and $\nu = 3.7,\ 2.0$ for \lcdm\, models with and without
tensor fluctuations respectively. For open CDM, without tensor
fluctuations, the powers are $\mu = 0.9$ and $\nu = 1.4$.  For the
span of models checked, the PS peak is in the range $0.02\leq k\leq
0.06\ihmpc$. The shape parameter of the $\Gamma$ model is only weakly
constrained to $\Gamma = 0.4\pm 0.2$. We caution, however, that these
results are as yet preliminary, and might depend on the accuracy of
the error estimates and on the exact impact of
non-linearities.\cite{fz}

\section*{Acknowledgments}
I thank my close collaborators in this work A.\ Dekel, W.\ Freudling,
Y.\ Hoffman and S.\ Zaroubi. In particular, I thank my collaborators
from the SFI collaboration, L.N.\ da Costa, W.\ Freudling, R.\
Giovanelli, M.\ Haynes, S.\ Salzer and G.\ Wegner, for the permission
to present these preliminary results in advance of publication.


\section*{References}
\begin{thebibliography}{99}

\bibitem{zz} S. Zaroubi, I. Zehavi, A. Dekel, Y. Hoffman and T. Kolatt,
\Journal{\apj}{486}{21}{1997}.

\bibitem{fz} W. Freudling, I. Zehavi, L.N. da Costa, A. Dekel, A. Eldar,
R. Giovanelli, M.P. Haynes, J.J. Salzer, G. Wegner, and S. Zaroubi, {\apj} 
submitted (1998). 

\bibitem{SW} M.A. Strauss and  J.A. Willick, 
\Journal{Phys. Rep.}{261}{271}{1995}.

\bibitem{k} N. Kaiser, \Journal{\mnras}{231}{149}{1988}.

\bibitem{mark3} J.A. Willick, S. Courteau, S.M. Faber, D. Burstein and 
A. Dekel, \Journal{\apj}{446}{12}{1995}; 
J.A. Willick, S. Courteau, S.M. Faber, D. Burstein, A. Dekel and
T. Kolatt, \Journal{\apj}{457}{460}{1996};
J.A. Willick, S. Courteau, S.M. Faber, D. Burstein, A. Dekel and
M.A. Strauss, \Journal{\apjs}{109}{333}{1997}.

\bibitem{sfi} R. Giovanelli, M.P. Haynes, L.N. da Costa, W. Freudling, 
J.J. Salzer and G. Wegner, in preparation.

\bibitem{sfiPOT} L.N. da Costa, W. Freudling, G. Wegner, R. Giovanelli, 
M.P. Haynes and J.J. Salzer,  \Journal{\apj}{468}{L5}{1996}.

\bibitem{ITF} L.N. da Costa, A. Nusser, W. Freudling, R. Giovanelli, 
M.P. Haynes, J.J. Salzer and G. Wegner, {\mnras} submitted (1997).

\bibitem{mockM3} T. Kolatt, A. Dekel, G. Ganon and J. Willick, 
\Journal{\apj}{458}{419}{1996}.

\bibitem{kd} T. Kolatt and A. Dekel, \Journal{\apj}{479}{592}{1997}.

\end{thebibliography}
\end{document}